\documentstyle [12pt,epsf]{article}

\newcommand{\beq}{\begin{equation}}

\newcommand{\eeq}{\end{equation}}
\newcommand{\bea}{\begin{eqnarray}}
\newcommand{\eea}{\end{eqnarray}}

\newcommand{\vf}{\varphi}

\newcommand{\e}{{\cal E}_\omega}

\begin{document}
\baselineskip 7 mm

\def\thefootnote{\fnsymbol{footnote}}

\begin{flushright}
\begin{tabular}{l}
CERN-TH/97-172 \\
hep-ph/9707423
\end{tabular}
\end{flushright}

\vspace{2mm}

\begin{center}

{\Large \bf 
New physics in a nutshell, or 
Q-ball as a power plant

}

\vspace{8mm}

\setcounter{footnote}{0}

Gia Dvali\footnote{ email address: dvali@mail.cern.ch},
\setcounter{footnote}{6}
Alexander Kusenko\footnote{ email address: kusenko@mail.cern.ch} 
and 
\setcounter{footnote}{1}
Mikhail Shaposhnikov\footnote{ email address: mshaposh@nxth04.cern.ch}
\\
Theory Division, CERN, CH-1211 Geneva 23, Switzerland \\

\vspace{12mm}

{\bf Abstract}
\end{center}

Future experiments may discover new scalar particles with 
global charges and couplings that allow for solitonic states.   If the
effective potential has flat directions, 
the scalar VEV inside a large Q-ball can exceed
the particle mass by  many orders of magnitude.  
Models with low-energy supersymmetry breaking generically have both the 
scalars carrying some global charges, and the flat directions. 
The Q-ball interior can, therefore, provide an environment for exploring
physics far beyond the TeV scale without the need for building  colliders
of ever-increasing energies.  Some Standard Model processes, otherwise
strongly suppressed,  can also be studied inside the soliton, where the
SU(2) symmetry can be restored, and the quark confinement may be absent.  
Baryon number violating processes catalyzed by the large VEV inside 
the Q-ball can provide  an inexhaustible energy resource.

\vspace{30mm}
 
\begin{flushleft}
\begin{tabular}{l}
CERN-TH/97-172 \\
July, 1997 
\end{tabular}
\end{flushleft}

\vfill

\pagestyle{empty}

\pagebreak

\pagestyle{plain}
\pagenumbering{arabic}
\renewcommand{\thefootnote}{\arabic{footnote}}
\setcounter{footnote}{0}

\pagestyle{plain}

\section*{Introduction}

The development of particle colliders has greatly advanced 
one's understanding of the fundamental laws of nature.  However, it is
becoming increasingly clear that growing costs and complexity of the 
accelerating technology will eventually stymie further exploration, unless 
some new paradigm will replace the concept of a collider experiment. 

In this Letter we point out that the new frontiers that will be unveiled by
the next generation of particle colliders, in particular the LHC, may allow
for a conceptually new approach to high-energy experimentation.  
A remarkable byproduct of such a development can be a new, practically 
unlimited, source of energy.  

New physics is essential for the applications we discuss.  First, we assume
that the new scalar particles, whose existence is anticipated
in most theories beyond the Standard Model, will be found in future
experiments.  Second, it is necessary that the low-energy effective
potential be invariant under a global U(1) symmetry.  Although the baryon
and lepton number conservation may suffice in principle, it is preferable 
for technical reasons (although not necessary) that there be a new U(1)
symmetry, with respect to which all the light fermions have zero charge.
Finally, the success of our idea relies on the appearance of ``flat
directions'' in the scalar 
potential, as is typically the case in theories with low-energy
supersymmetry.   We will assume hereafter that the new physics at the
electroweak scale meets all three conditions.  We emphasize that
supersymmetry {\it per se} is not essential for our idea, which is quite
general, but the presence of scalar fields and flat directions makes
supersymmetric models a natural setting for applications discussed below.

Our basic idea is that the manifestations of the high-scale physics
can be attained in the interior of a relatively long-lived artificially
created non-topological soliton of the Q-ball type \cite{tdlee,coleman1}.   
A specific property of such solitons is that the scalar VEV $\langle \phi
\rangle$ inside a large Q-ball tends to take the value which minimizes the
ratio of the interior energy density to $\langle \phi^2 \rangle $, that is 
$ U(\phi)/\phi^2 = min$ \cite{coleman1}. 
If the potential $U(\phi)$ has flat directions, as expected in theories
with low-energy supersymmetry, then the scalar VEV inside a sufficiently 
large Q-ball can exceed the energy density by many orders of magnitude.  
Various couplings, suppressed by the powers of $\langle \phi \rangle / 
M_{new \ physics} \ll 1 $ in the low-energy effective Lagrangian, become
large inside the Q-ball and can mediate exotic processes,  
whose detection can yield information about the new physics.  

In addition, the ground state of matter inside a large Q-ball is different
from the standard vacuum.  Various processes, allowed by the Standard
Model but suppressed in the vacuum with a broken electroweak symmetry, 
can take place in the soliton interior where the SU(2) can be restored, 
and the quark and gluons may not be confined.

A particular example of an exotic process predicted by Grand 
Unified and other theories, is the explicit baryon number violation.  
A Q-ball with a baryon number violating interior can be used to generate 
energy through a decay of the incoming nucleons into leptons and photons.

\section{Far-reaching Q-balls} 

In this section we study non-topological solitons \cite{tdlee,coleman1} 
in the potential $U(\vf)$ that is essentially flat for large values of
$\vf$, by which we mean that it grows slower than $\vf^2$ (Figure~1).  
As discussed in the next section, such potentials are common in theories
with softly-broken low-energy supersymmetry, where numerous flat directions
are lifted only by the soft supersymmetry breaking terms and Planck scale 
suppressed operators. Although supersymmetry is not essential for what
follows, it provides a natural motivation for an otherwise {\it ad hoc} 
potential we employ.

\begin{figure}
\setlength{\epsfxsize}{3.3in}
\centerline{\epsfbox{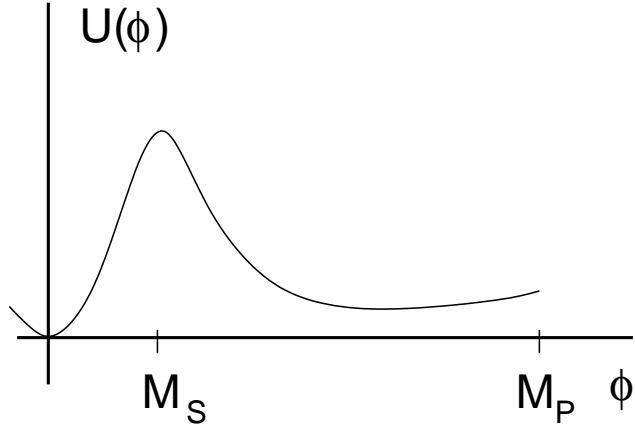}}
\caption{
A potential with a flat direction lifted only by logarithmic corrections 
at large VEV. 
}
\label{fig1}
\end{figure}

If $U(\vf)$ is invariant under a global U(1) symmetry, $\vf \rightarrow
\exp(i\theta) \vf $, then the theory admits non-topological solitons, 
Q-balls, which are the minima of energy in the sector of fixed charge. 
Such solitons can be described as time-dependent spherically symmetric
lumps $\vf(x,t) =e^{i \omega t}  \vf(r), \ r^2=\vec{x}^2$, where the 
function $\vf(r)$ is such that it extremizes the functional \cite{ak_qb} 

\beq
\e = \int d^3x \, \left [\frac{1}{2} |\nabla \vf(x) |^2  
+ \hat{U}_\omega(\vf(x))
\right ] + \omega Q .
\label{Ew}
\eeq
Here $Q= \frac{1}{2i} \int \vf^* \stackrel{\leftrightarrow}{\partial}_t  
\vf \, d^3x $ is the soliton charge and $\hat{U}_\omega(\vf)
=U(\vf)-(\omega^2/2) \vf^2$.   The equation of motion for $\vf(r)$ is that
of a bounce associated with tunneling in three Euclidean dimensions in the
potential $\hat{U}_\omega(\vf)$: 

\beq
\left \{ 
\begin{array}{l}
\vf''(r)+(2/r) \vf'(r) = \partial \hat{U}_\omega /\partial \vf \\ \\
\vf(0)=\vf'(\infty)=0.
\end{array}
\right.
\label{eqn_mtn}
\eeq

In the thin-wall approximation (whose applicability will be 
discussed below), Q-ball is approximated by an ansatz 

\beq
\vf(r) = \left \{ \begin{array}{ll}
                    \vf_0, \  & r < R \\ & \\
                    0, \  & r \ge R
                 \end{array}
         \right.
\label{thinwall}
\eeq
Both $\omega $ and $\vf_0$ are found by minimizing the energy of the
soliton. The value of the scalar field in the Q-ball interior is such that 

\beq
\frac{U(\vf)}{\vf^2} = min \ \ \ {\rm for} \ \ \ \vf=\vf_0
\label{min}
\eeq

Of course, for a potential of the kind shown in Figure~1, which grows
slower than $\vf^2$ for large $\vf$, the thin-wall Q-ball would appear to
have an infinite VEV.  In fact,  condition (\ref{min}) determines the value
of $\vf_0$ only in the thin-wall limit, that is, when the gradient energy is
small in comparison to the volume energy, and the ansatz (\ref{thinwall}) is 
appropriate.  As $\vf(0)$ increases, the thin-wall approximation breaks
down for a Q-ball of a fixed charge.  For a flat potential, the equation of
motion that determines the shape of the soliton  can be
solved analytically near the origin $r=0$, where the right-hand side of the 
differential equation (\ref{eqn_mtn}) is simply $\omega^2 \vf$.  
For large $r$, $\vf(r) \sim \exp \{-m_\vf r\}$, where $m$ is the mass of 
$\vf$ near the origin.  The solution can be approximated by 

\beq
\vf(r) = \left \{ 
\begin{array}{ll}
\vf_0 (\sin \omega r)/\omega r, \ & r < R \\  & \\ 
\vf_1 \exp \{-m_\vf r\}, \ & r \ge R,  
\end{array}
\right.
\label{ansatz}
\eeq
where the values of $\vf_0$, $\vf_1$, $\omega$  and $R$ 
are such that they minimize $\e$ in equation (\ref{Ew}), while 
$\vf(r)$ is continuous at $r=R$.

Clearly, the VEV of $\vf$ slides along the flat direction, in accord with 
(\ref{min}), until the solution is way outside the thin-wall regime.  In
this thick-wall limit, one can rewrite $\e$ in equation (\ref{Ew}) in terms
of dimensionless variables $\xi = \omega x$ and $\psi = \vf/\omega $
neglecting all terms in $\hat{U}_\omega $ except the constant term and the 
$-(\omega^2/2) \vf^2$ term ({\it cf}. Refs. \cite{ak_qb,linde}).  
The resulting expression for $\e$ is 
$\e \approx a \omega + b/\omega^3 + \omega Q$, where $a$ and $b$ are
constants independent of $\omega$.  The value of $\omega$ that minimizes
$\e$ depends on the charge as $\omega \propto Q^{-1/4}$.  The size of the 
soliton (\ref{ansatz}) is $R \sim 1/\omega$.  Therefore, the value of the
field in the Q-ball interior 

\beq
\vf_0 \propto Q^{1/4}.
\label{vev}
\eeq

We conclude that the scalar VEV inside a Q-ball can exceed the scales
$M_{_S}$ and $m_\vf$ by many orders of magnitude, provided that the charge of
a soliton is large.   If the $\vf$ particles can be produced in a collider,
then, conceivably, one can build a large Q-ball by accumulating the U(1)
charge.  As the VEV of $\vf$ slides along a flat direction, 
the interior of the soliton can provide access to new physics at 
the scales $\vf_0 \sim Q^{1/4} m_\vf \gg m_\vf$ for large $Q$.   Of course,
it would require an enormous charge to reach, for example, the GUT scale. 
The requisite Q-ball would have to carry charge as large as $10^{56}$ and 
its mass would be of order $10^{20}$ g.   However, conceivably, one can 
hope to attain the intermediate scales $\sim 10^{8}$ GeV using the solitons 
with charges of order $ N_{_A}\sim 10^{23}$. 

Another important property of a large Q-ball in a flat potential is that
its energy grows as $E \propto Q^{3/4} $, rather than $Q$, which means that
storing charge in the maximal size solitons is energetically preferable.  
This is because the Q-ball in a flat potential never
becomes a thin-wall object.  If the scalar VEV remained constant, the 
soliton mass would grow as the first power of charge.  However, in our case, 
the VEV of $\vf$ slides out along the flat direction as the charge
increases, thus accommodating more charge at the lesser energy expense than
it would be in the thin-wall limit.  Hence, $E$ grows slower than $Q$. 

\section{Flat potentials in theories with broken supersymmetry}

The effective potentials of the type specified in the previous section 
can be naturally realized (and are, in
fact, generic) in theories with supersymmetry-breaking soft masses 
originating at some low energy scale $\ll M_P$, for example, in 
models with gauge-mediated supersymmetry breaking~\cite{gauge-mediated}.
To avoid the problematic supertrace relation, 
it is commonly assumed that the supersymmetry breaking takes place 
in some hidden sector, that is, the sector that has no
direct couplings to the quark and lepton superfields in the superpotential.
The role of this sector is to provide a superfield(s) $X$ (usually a singlets
under the standard model group) with a nonvanishing scalar and auxiliary
($F_X$) components. This breaks supersymmetry, and also ensures that no
unbroken $R$-symmetry survives. 
 The transmission of the supersymmetry breaking to the observable sector
is due to some messenger interaction with a typical scale $M_{_S}$. 
Supergravity, or some heavy particles charged under the
standard model gauge group, can be the messengers in the so called 
gravity-mediated or gauge-mediated scenarios, respectively. 
Integrating out the messenger sector below the scale $M_{_S}$, one is left
with the higher dimensional couplings (suppressed by powers of
$M_{_S}^{-1}$) between the observable and the hidden sector
superfields. The resulting scale of supersymmetry breaking in the
observable sector is set by the ratio $F_X/M_{_S}$. In this scenario the
soft masses are "hard" below the  scale $M_{_S}$ and disappear above that
scale. Thus, in the gravity-mediated  scenarios these soft masses stay
intact, up to the renormalization effects,  all the way up to the Planckian
energies (or the field strengths),  whereas in gauge-mediated scenarios
with low $M_{_S}$ they die-off  above the scale $M_{_S} << M_P$ . For this
reason, the behavior of the scalar potentials differs dramatically in these
two cases.   In our analysis we will concentrate on the effective
potentials of the flat direction fields, which are very generic in
supersymmetric theories. 

 It is well known that the MSSM and its extensions admit a variety of 
non-compact flat vacuum directions in the unbroken supersymmetry
limit. There are certain combinations of the scalar fields (squarks and
sleptons), such that $D$ and $F$ terms in the  scalar potential 
are identically zero for  
arbitrary expectation values.  Each flat direction can be parameterized
by some holomorphic invariant constructed out of the chiral superfields
$I = Q_1Q_2...Q_N$. Let $\vf =  I ^{1/N}$  be a canonically
normalized flat direction field. Generically, $Q_i$ are charged under
the standard model gauge group. Along the flat direction, 
all the gauge and chiral fields coupled to $\vf$ get masses $\sim \vf$ and
become heavy. As the VEV becomes large, one can integrate them out and
write an effective low energy theory. It will consist of the massless
chiral superfield $\vf$ which is essentially decoupled from the rest of the
light particles.  The rest of the interactions affect the massless field 
$\vf$ only through some higher-dimensional operators suppressed by $\langle
\vf \rangle$. 
In addition, the supergravity corrections induce similar operators
suppressed by $M_{_P}$, which become important for $\langle \vf \rangle 
\sim M_P$.  Thus it is not surprising that in theories with low-energy
supersymmetry breaking the potential becomes essentially flat
for $\vf \gg M$, up to the corrections due to gravity, which can lift it
for very large field values.

\begin{figure}
\setlength{\epsfxsize}{2.0in}
\setlength{\epsfysize}{2.3in}
\centerline{\epsfbox{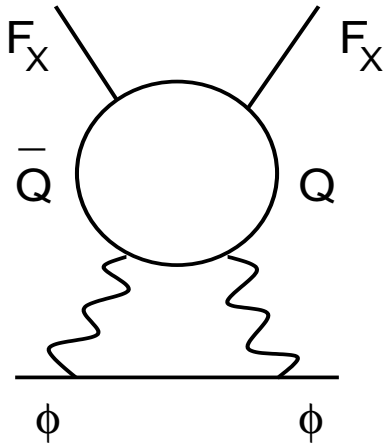}}
\caption{
Supersymmetry breaking in the hidden sector can be  communicated to 
{$\vf$} via messengers and gauge interactions. 
}
\label{fig2}
\end{figure}

For example, let us consider the case of gauge-mediated supersymmetry
breaking.  In this case the supersymmetry breaking is communicated to the
observable sector, which includes $\vf$ and other fields, by gauge
interactions via a diagram shown in Figure~2.  Here $Q$ and 
$\bar{Q}$ are the messengers, chiral superfields that share some common
gauge interaction with $\vf$.  The corresponding gauge superfield is
propagating in the lower loop.  By assumption, there is also a tree-level 
coupling of messengers to the $X$ superfield in the superpotential, $X
\bar{Q} Q$. They acquire a supersymmetric mass $\sim \langle X
\rangle$ from this  coupling.
At the low energy scales, or, equivalently, for $\vf < \langle X
\rangle^{1/2}$,  the flat potential for $\vf$ is lifted by a soft mass
term $\sim  \left ({\alpha \over 4\pi} \right ) (\langle
F_{_X} \rangle/ \langle X \rangle)$, the gauge fields in Figure~2 being
light. However, farther out along the $\vf$ direction, the VEV
of $\vf$ induces 
a mass of order $\langle \vf \rangle $ for the gauge superfields in the
two-loop diagram (Figure~2), which becomes dominated by the $1/ \langle
\vf \rangle^2 $ factors from integrating over the gauge fields momenta 
in the lower loop.  Therefore, at the scales (field strengths)
much larger than $\langle
X \rangle^{1/2}$, the soft mass of the $\vf$ field gradually
turns off, and the effective potential approaches zero asymptotically. 
Eventually, the Planck-scale corrections are expected to modify $U(\vf)$ 
in the limit $\langle \vf \rangle \rightarrow M_{_P}$ and lift the
the potential above zero. In general, there can be a variety of
other minima in this region,  which we will ignore. In any
case, a qualitative behavior of the classically-flat potential 
in models with gauge-mediated (or {\it any other} low-energy) supersymmetry
breaking is as shown in Figure~1.

\section{Toy model}

As a toy model, let's consider a theory with a scalar $\vf$  and two
light fermions, $q$ and $l$, with different masses $m_q > m_l$.  The
relevant part of the Lagrangian is   
\beq
V=U(\vf) + y \frac{\vf^\dag \vf}{M_{_X}} (\bar{q}l) + h. c. , 
\label{toy}
\eeq
where the potential $U(\vf)$, of the kind specified in the previous
sections and shown in Figure~1, is augmented by an effective coupling 
to fermions suppressed by the 
large mass scale $M_{_X} \gg m_{\vf}$ (such non-renormalizable interaction 
can arise naturally in realistic theories, as shown below).  
By assumption, $U(\vf) $ is invariant under a global U$_\vf$(1) symmetry 
$\vf \rightarrow e^{i \theta} \vf$. In addition, for $y=0$ there  is a
``baryon'' U$_{_B}$(1) symmetry $q \rightarrow e^{i \alpha} q$, and an
analogous ``lepton'' number conservation for $l\rightarrow e^{i \beta} l$.  
In vacuum, $\langle \vf \rangle =0 $ and the transition of $q$ into 
$l$, our toy version of ``proton decay'', is disallowed at tree level.  
The first contribution to the $q \rightarrow \bar{l}ll $ decay arises at the
two-loop level and the rate is small: $\Gamma \sim (y m_\vf / M_{_X})^6
m_q$.  

Due to the U$_\vf$(1) symmetry, the theory
admits stable Q-balls.  Inside a Q-ball, a non-zero VEV of $\vf$ enhances 
the probability of the $q \rightarrow l$ transition\footnote{
We note in passing that the momentum conservation in such decay is
accommodated by the Q-ball recoil: $q \rightarrow l + {\rm sound \ waves}$
(see discussion in Ref. \cite{coleman1}).} and the heavy
fermion, q, has the decay width 

\beq
\Gamma \sim y^2 \frac{\langle \vf \rangle ^2}{M_{_X}^2} m_q
\eeq
that can be large for a sufficiently large VEV of $\vf$.
One can, therefore, use the Q-ball interior to catalyze the ``proton decay''
$q \rightarrow l$ and utilize the energy released in this process.   In
addition, because of the $q-l$ mixing induced by the coupling of $q$ and
$l$ to $\vf$, a $q$-particle scattering off a large Q-ball can be reflected
as an $l$-particle with a probability 

\beq
P\sim  y^2 \frac{\langle \vf \rangle ^4}{m_q^2 M_{_X}^2}, 
\eeq
thus providing conditions for converting incoming ``protons'' into
``leptons''.

As a variation on this model, one can consider a Yukawa interaction
\beq
V=U(\vf) + y \vf (\bar{q}l) + h. c. , 
\label{toy1}
\eeq
where the Yukawa coupling $y$ is very small.  Now the U$_\vf$(1) symmetry
is approximate: it is broken by the small Yukawa coupling.  The
corresponding Q-ball is as stable as the symmetry is good and can slowly 
evaporate by emitting light fermions.

\section{The MSSM and its extensions}

The assumptions we made about the particle content of the theory and the
form of the effective potential can be naturally realized in theories with
softly broken supersymmetry.   
To create a Q-ball useful for probing the new physics, one needs a complex 
scalar field that transforms under a global U(1) symmetry.  If a large 
VEV of such scalar can facilitate a proton decay, then the same 
Q-ball will also be useful for the purposes of generating power 
by releasing the energy locked in the nucleons by the 
baryon number conservation.  

In the MSSM, the supersymmetric partners of the quarks and leptons can form
Q-balls whose stability (neglecting the effect of fermions) owes to the
conservation of the baryon and lepton 
numbers \cite{ak_mssm}.  Under certain conditions, such objects can be
created in the early Universe and can precipitate a decay of a false
vacuum even if the tunneling probability is negligible \cite{ak_pt}.
Squarks and sleptons are, therefore, natural candidates for 
creating a Q-ball in a laboratory.  
However, in the presence of the light fermions that carry the
same charge, these solitons can evaporate by emitting quarks and leptons
from the surface as discussed in Ref. \cite{ccgm}.  The rate of such
evaporation is proportional to the surface area, rather than the volume of
a Q-ball, and can only be neglected if a sufficiently large Q-ball can be
built.  While not fatal in principle, this can be seen as a considerable
complication.   

To illustrate the general idea, we turn to the case in which the
issue of stability is simplified by the absence of light 
U(1)-charged particles in the theory,  and where 
the ``flat'' potential is realized in terms of colorless degrees of freedom.  
The considerations of stability are trivial if an additional scalar field,
whose couplings to matter fermions are small, is used to create a soliton.
Such fields, in fact, do appear in  various extensions of the Standard
Model. 

An example of the light decoupled scalars can be provided by the
color-triplet Higgs ($T, \bar T$) 
partners of the electroweak Higgs doublets. These particles are inevitably
present in any GUT extension of the MSSM and carry a non-zero baryon and
lepton number, since they are coupled to the ordinary quarks and leptons.
To explain the proton stability, one usually assumes these particles 
to be very heavy.
This is the well known doublet-triplet splitting problem in SUSY GUTs.
The enormous mass splitting, however, can be avoided 
if the triplets are light but have very weak, GUT scale suppressed 
interactions with quarks and leptons~\cite{gion}.
Such suppression is due to Clebsch factors which are determined by the 
group structure of the the GUT-breaking VEVs.
This can be achieved naturally, 
as in the model of Ref. \cite{gion}.  
Since the Higgs triplets are essentially decoupled
(from quarks and leptons), the low energy action below the GUT scale
contains an additional unbroken global symmetry 
G$=U'_T(1)\times U''_{_T}(1)$ which transforms $T \rightarrow e^{i
\theta} T$ and $\bar{T} \rightarrow e^{i \chi} \bar{T}$. 
This global symmetry is only broken by small couplings,   
suppressed by the powers
of the GUT scale.  In a theory with gauge-mediated supersymmetry breaking 
the triplets have soft masses of order $\sim 1$~TeV which vanish at 
energy scales large in comparison to the messenger mass.  The corresponding
potential has a flat direction for some combination $\vf$ of $T$ and
$\bar{T}$, charged under a subgroup  U$_T$(1) of G. 
This potential satisfies the conditions discussed in the previous
section and is suitable for building U$_T$(1) Q-balls. 

Depending on the cosmological history of the Universe, the triplets may or
may not be subject to constraints related to the non-observation of the 
``wild hydrogen'', and other exotic elements whose nuclei might contain a
heavy stable colored $T$-particle.  If inflation took place and ended
with a relatively low reheating temperature, sufficiently below the $T$
mass, then  no constraints on the mass and lifetime of the triplets arise.
Otherwise,  $T$ particles must decay in less than $10^{10}$ years to evade
the superheavy element searches and other bounds \cite{cosm}.
Different considerations \cite{derujula} exclude the (electrically charged)
stable $T$ particles outside the 20 to 1000 TeV window.  In any case, even
without the help of a relatively cool inflation, weakly-interacting $T$
particles with lifetimes $>10$ s and masses in the TeV region are allowed 
by the data. 

If the color triplets are discovered, they can be pair-produced in a
hadron collider.  If stable, the $T$ particles can also be found on Earth, 
although their concentration is constrained to be small \cite{cosm}.
We assume that the $T$-particles can be condensed into Q-balls which 
will grow as one feeds in some additional charge.  Technical and
engineering aspects of such process, which may or may not be possible to
realize in practice,  lie outside the scope of our investigation. 

There are two obvious candidates for building stable Q-balls in this model.
We assume that the soft masses for both the squark $q$ and the $T$-field
originate at the scale $M_{_S}$ and effectively turn off 
at higher energy scales, as explained in section 2.  The effective
potential in terms of each of the colorless fields $\vf_1 = (T^a\tilde
q^a)^{1/2}$ and $\vf_2= (T^a \bar{T}^a)^{1/2}$ is
then of the form specified above, with a nearly flat direction.  Both 
$\vf_1$ and $\vf_2$ are suitable for Q-balls.  Below we will often 
use $\vf$ for either of the two fields. 
The field $\vf$ inside the Q-ball develops a large VEV limited
only  by the Q-ball size. The increase of a scalar VEV inside a growing
Q-ball can eventually trigger some exotic processes characteristic of the
new physics at the high energy scale.  Presumably, it would take only 
a TeV-scale machine to produce the $T$ particles.  However, if a large
enough Q-ball is created, one can explore the new frontiers well above the
TeV energy scale.

If the scalar field that makes up a Q-ball can decay into quarks, then 
there is a finite density of quarks (of order $m_\vf^3$) inside the Q-ball.
The interior of such Q-ball is similar in nature to the cold but dense
quark-gluon plasma.  This is not the case if $\vf$ particles are stable. 
If the scalar VEV is sufficiently large, the electroweak symmetry may be
restored inside the Q-ball.  For the SU(2)-nonsinglet fields $\vf$ 
the running of couplings is frozen at the scale $\mu \sim \vf$, which may
correspond to an unbroken SU(2)$\times$U(1). 

For example, 
in theories with gauge-mediated supersymmetry breaking, the soft mass 
of the Higgs boson originates below the scale $M_{_S}$ at the two-loop
level:  

\beq
m_{_H}^2= \mu^2 + \left [ 
c_1 \left  ( \frac{\alpha_1}{4\pi} \right )^2+
c_2 \left  ( \frac{\alpha_2}{4\pi} \right )^2 \right ]
\frac{F_{_X}^2}{M_{_S}^2} 
\label{higgsmass}
\eeq
where $\alpha_i$ are the gauge couplings
and $c_1$ and $c_2$ are the group factors.  The negative contribution to
the Higgs mass, necessary for the electroweak symmetry breaking, comes at
the three-loop level due to the strong Yukawa interaction with the top: 

\beq
\Delta m_{_H}^2= - \frac{6 h_t^2}{16\pi^2} 
\ln \left ( \frac{\Lambda}{\tilde{m}_t} \right ) \tilde{m}_t^2
\eeq
where $h_t$ is the top Yukawa coupling, and $\tilde{m}_t^2 \sim \left (
\frac{\alpha_3}{4\pi} \right )^2  \frac{F_{_X}^2}{M_{_S}^2}$ is the
two-loop mass of the stop.  
This contribution can be dominant, thanks to the large value of $\alpha_3$.  

If the VEV's of $T$ and
$\tilde{q}$ are large inside the Q-ball built of $T$-fields and the 
right-handed (SU(2)-singlet) squarks, then $\alpha_3$ is small. 
The negative contribution in equation
(\ref{higgsmass})  diminishes and $m_{_H}^2$ can become positive in the
Q-ball interior.  Thus, the electroweak symmetry is restored and, in
particular,  the sphaleron transitions can go unsuppressed.  In this sense
the ground state inside the Q-ball is similar to that of the electroweak
bags~\cite{khsh}.   

We now want to examine the feasibility of using Q-balls for power
generation via the nucleon decay.  There are several processes that can 
yield large amounts of energy once a Q-ball is assembled and placed in a 
beam of protons.   

First, the dissociation of protons into quarks can yield $\sim 1$ GeV of
energy per nucleon.  For a large VEV of $\vf$ the 
QCD coupling becomes small. 
As a result, an incoming proton will dissociate into quarks and
dissipate  the binding energy $\sim 1$ GeV in photons and ``sound waves'' 
\cite{coleman1}.  These excitations can propagate to the Q-ball surface, 
where the energy is emitted in photons and pions pair-produced near the
soliton boundary.  

Second effect is the sphaleron transitions of quarks into leptons, which
are possible if the SU(2)$_{_L}\times$U(1) symmetry is restored inside the
Q-ball, much like in the case of the electroweak bags~\cite{khsh}.   

Third, as in the toy model of section 3, the incoming protons can scatter
off the Q-ball boundary as leptons.  The $T$ VEV can induce a mixing
between quarks and leptons  \cite{gion}.  If $\langle T \rangle$ is large,
the mass eigenvalues for quarks and leptons inside the Q-ball can be larger
than those in vacuum.  A soliton boundary can then serve as a repulsive
potential barrier for protons.  Due to the large quark-lepton mixing in the
barrier, the back-scattered fermions will be leptons with non-vanishing 
probability.   
 
Proton decay catalyzed by the presence of the Grand Unified magnetic
monopoles \cite{cr} has been considered as a potential source of energy. 
It was estimated \cite{beno} that $10^{11}$ monopoles would be needed to
replace one industrial power plant.  In the case of Q-balls, the efficiency
of power generation would depend on their size.  For instance, a single  
Q-ball placed in water would catalyze proton decay at the rate proportional
to its surface area.  An ``industrial'' amount of power $\sim 10^8$ W 
could be achieved with the charge $10^{40}$ Q-ball, a few microns in
diameter, weighing 100 tons.  The same efficiency can be achieved 
using $10^{17}$ Q-balls with charge $10^{6}$ each, hence reducing the total
charge to $10^{23}$.   Although the surface-to-volume ratio favors a higher
number of smaller Q-balls, there is a model-dependent limit on the charge
of the smallest soliton that catalyzes a proton decay in one of the way
discussed above.

\section*{Conclusion}

Scalar particles, such as squarks, sleptons, and others, 
may be discovered at the scales $\stackrel{<}{_{\scriptstyle \sim}}
1$~TeV, which will become
accessible to particle accelerators in the near future.  The new scalar
fields may transform non-trivially under $B$, $L$, or some new global
symmetries and may have the requisite couplings \cite{ak_mssm}  suitable 
for Q-balls.  

If, as expected in SUSY theories, the scalar potential has flat 
directions, then the Q-ball interior may provide the environment for 
exploring physics far beyond the TeV scale without the need for building 
colliders of such enormous energies. Instead, one can learn to create large
Q-balls in a laboratory setting, and use them as instruments for high-energy
experimentation.  The exotic processes that can take place inside a large
artificially-built Q-ball can be detected and analyzed, thus helping
elucidate the new physics. 

Normally suppressed, baryon number violating processes may become unleashed
inside the Q-ball.  A controlled proton decay catalyzed by a Q-ball can
serve as an inexhaustible source of energy.  
Although speculative, this idea requires few assumptions about 
the physics beyond the Standard Model: we have assumed the MSSM with
low-energy supersymmetry breaking.  As was explained earlier, some
additional scalars with new conserved U(1) quantum numbers can be of help,
but are not necessary.

We thank A.~Cohen, S.~Coleman and P.~Tinyakov for helpful discussions.

\end{document}